\documentclass[twocolumn,showpacs,superscriptaddress,preprintnumbers]{revtex4}
\usepackage{amssymb}
\usepackage{graphicx}
\usepackage{dcolumn}
\usepackage{bm}

\begin{document}

\title{An effective thermal-parametrization theory for the slow-light
dynamics in a Doppler-broadened electomagnetically induced transparency
medium}
\date{Feb 25, 2011}
\author{Shih-Wei Su}
\affiliation{Department of Physics, National Tsing Hua University, Hsinchu 30013, Taiwan}
\author{Yi-Hsin Chen}
\affiliation{Department of Physics, National Tsing Hua University, Hsinchu 30013, Taiwan}
\author{Shih-Chuan Gou}
\email{scgou@cc.ncue.edu.tw}
\affiliation{Department of Physics, National Changhua University of Education, Changhua
50058, Taiwan}
\author{Ite A. Yu}
\email{yu@phys.nthu.edu.tw}
\affiliation{Department of Physics, National Tsing Hua University, Hsinchu 30013, Taiwan}

\begin{abstract}
We model the effects of atomic thermal motion on the propagation of a light
pulse in an electromagnetically induced transparency medium by introducing a
set of effectively temperature-dependent parameters, including the Rabi
frequency of the coupling field, optical density and relaxation rate of the
ground state coherence, into the governing equations. The validity of this
effective theory is verified by the close agreement between the theoretical
results and the experimental data.
\end{abstract}

\maketitle

\address{$^{1}$Department of Appled Mathematics, Feng Chia University, Taichung 40724,\\
Taiwan\\
$^{2}$Department of Physics, National Changhua University of Education,\\
Changhua 50058, Taiwan\\
$^{3}$Department of Mathematics, National Taiwan University, Taipei 106,\\
Taiwan}


\section{INTRODUCTION}

The ability of controlling light is crucial to the fulfillment of quantum
communications and all their practical applications. It has been
demonstrated recently that a light pulse can be slowed down \cite%
{Boller,Field} and even stored in media \cite{Phillips,Kocharovskaya} by
using the effect of electromagnetically induced transparency (EIT), a
nonlinear optical phenomenon that renders an opaque medium transparent by
irradiating it with an electromagnetic field \cite{M. Fleischhauer} .
Subsequent studies indicate that the slow light (SL) effect can greatly
enhance optical nonlinearity at the low light level \cite{Harris2} , and has
potential applications in the development of low-light-level and
single-photon devices, such as the all-optical switches \cite{Yamamoto} and
quantum phase gates based on the cross-phase modulation scheme \cite%
{Schmidt,Imamoglu} . The novelty of EIT thus provides a new fashion to
manipulate the behavior of light, and has attracted a great deal of research
interest.

\begin{figure}[htbp]\begin{center}
\includegraphics[width=1.6734in]{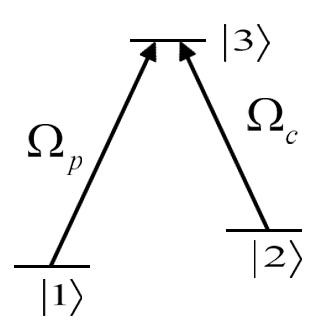}
\caption{(Color online) The energy levels of a 3-level atom of $\Lambda $-type. The transitions $
\left\vert 1\right\rangle \leftrightarrow $ $\left\vert 3\right\rangle $ and
$\left\vert 2\right\rangle \leftrightarrow $ $\left\vert 3\right\rangle $
are driven by laser fields with Rabi frequencies $\Omega _{p}$ and $\Omega
_{c}$ correspondingly.}
\label{sag}
\end{center}\end{figure}

The prototype of EIT systems consists of a weak probe beam of central
frequency $\omega _{p}$ and a strong coupling beam of frequency $\omega _{c}$
interacting with a three-level atom, as shown in Fig.1. The dynamics of the
slowly varying density matrix elements in the weak probe limit are described
by the optical Bloch equation \cite{Zubairy}

\begin{equation}
\frac{\partial \rho _{12}}{\partial t}=-\left[ i\left( \Delta _{p}-\Delta
_{c}\right) +\gamma \right] \rho _{12}-\frac{i\Omega _{c}}{2}\rho _{13}
\label{Bloch12}
\end{equation}

\begin{equation}
\frac{\partial \rho _{13}}{\partial t}=-\left( i\Delta _{p}+\frac{\Gamma }{2}%
\right) \rho _{13}-\frac{i\Omega _{c}^{\ast }}{2}\rho _{12}-\frac{i\Omega
_{p}^{\ast }}{2}  \label{Bloch13}
\end{equation}%
where $\Omega _{p}$ and $\Omega _{c}$ denote the Rabi frequencies driving
the transitions $\left\vert 1\right\rangle \leftrightarrow \left\vert
3\right\rangle $ and $\left\vert 2\right\rangle \leftrightarrow \left\vert
3\right\rangle $ respectively; $\gamma $ is the relaxation rate of the
ground-state coherence, and $\Gamma $ is the spontaneous decay rate of the
excited state $\left\vert 3\right\rangle $. The detunings $\Delta _{p}$ and $%
\Delta _{c}$ are defined by $\Delta _{p}=\omega _{31}-\omega _{p}$ and $%
\Delta _{c}=\omega _{32}-\omega _{c}$, where $\omega _{ij}$ denotes the
transition frequency between the energy levels $\left\vert i\right\rangle $
and $\left\vert j\right\rangle $. The propagation of the probe pulse is
described by the Maxwell-Schr\"{o}dinger equation

\begin{equation}
\frac{1}{c}\frac{\partial \Omega _{p}\left( z,t\right) }{\partial t}+\frac{%
\partial \Omega _{p}\left( z,t\right) }{\partial z}=i\eta \rho _{31}
\label{Maxwell}
\end{equation}%
where $\eta $ measures the opacity per unit length of the medium and is
related to the optical density of the medium by $d^{opt}=2L\eta /\Gamma $ ($L
$ is the medium length). Here $\rho _{31}$ corresponds to the slowly varying
amplitude of the optical coherence of the probe transition.

We note that a decay constant $\gamma $ is introduced in eq.($\ref{Bloch12}$)
to account for the relaxation of $\rho _{21}$.\textbf{\ }Phenomenologically,
we can write $\gamma =\gamma _{T}+\gamma _{0}$, where $\gamma _{T}$
represents the effects originating from the thermal random motion of atoms
and $\gamma _{0}$ represents other intrinsic effects, such as the linewidth
and frequency fluctuations of the laser fields, the stray magnetic field,
and the collisions between atoms\textbf{, }etc. Except the demonstration of
SL in Bose-Einstein condensate \cite{Hau,Liu} , most EIT experiments have so
far been carried out using either laser-cooled atoms or hot atoms, with
temperatures widely ranging from milliKelvins to room temperature \cite%
{Budker,Kash} . It is recognized that the randomized Doppler shifts stemming
from the thermal motion of atoms is the predominating mechanism of
decoherence at finite temperatures. Since the temperature as an independent
variable does not enter eqs.($\ref{Bloch12})-(\ref{Maxwell}$), $\gamma _{T}$
is bound to depend implicitly on the temperature and can only be determined
by numerical fitting with the experimental data.

In order to gain more insight into the dynamical properties of an EIT medium
at finite temperatures, it is desirable to take the atom's motion into
account by statistical approach. Javan \textit{et. al.} suggested that the
macroscopic polarization, which serves as a response to the randomized
frequency shift, should be obtained by taking average over all possible
velocities characterized by the Maxwell-Boltzmann distribution \cite%
{Scully,Zhao} . It is expected that such an averaging process smears the
atomic coherence and thus contribute largely to the relaxation of $\rho
_{12} $. More recently, the dynamics of SL, light storage and stationary
light pulse in finite-temperature EIT media are studied in \cite{Su,Wu} . Wu
\textit{et. al. }\cite{Wu}\textit{\ }have developed another method to solve
the dynamics of stationary light pulses in a cold atomic cloud by solving
the Maxwell-Liouville equations which contain the higher order spin and
optical coherence and the Doppler effect arising from the atomic thermal
motion. By using the gauge-invariance of Schr\"{o}dinger equations under the
Galilean transformation, Su \textit{et. al.} \cite{Su} have developed a
numerical scheme to study the dynamics of SL and light storage in a
Doppler-broadened EIT medium, in which the effects of the atom's external
degrees of freedom are fully considered.

In this paper, we investigate the dynamics of SL in a Doppler-broadened EIT
based on the formalism of optical Bloch equations. The central idea of this
work is to treat the atom as a stationary object but otherwise introduce
some temperature-dependent parameters which are responsible for the
manifestation of thermal effects. Specifically, we analyze the output probe
pulse under the influence of the atomic thermal motion by endowing the
coupling field, the optical density and the phenomenological decay constant
with temperature dependence, namely, we let $\Omega _{c}\rightarrow \bar{%
\Omega}_{c}\left( T\right) $, $\eta \rightarrow \bar{\eta}\left( T\right) $,
$\gamma \rightarrow \bar{\gamma}\left( T\right) $ for a stationary EIT
medium. Derivation of these effective parameters will be described in the
following.

\section{FORMALISM}

To begin with, we assume that the shape distortion of the output probe pulse
can be treated as a superposition of plane waves of different frequencies
propagating through the highly dispersive and slightly absorptive EIT medium
with length $L$. Each output plane wave attains a factor of $e^{i2\pi
L(n_{R}+in_{I})/\lambda }$ while propagating inside the medium, where $n_{I}$
and $n_{R}$ are the real and imaginary parts of the refractive index of the
medium, which lead to the attenuation and phase shift of the output plane
waves, respectively. Note that the refractive index is related to $\rho _{31}
$ by $2\pi \left( n_{R}+in_{I}\right) /\eta \lambda =\rho _{31}/\Omega _{p}$
\cite{Kao} .

In the weak probe limit, the steady state solution of $\rho _{31}$ of the $%
\Lambda $-type EIT system is given by \cite{Kao}

\begin{equation}
\frac{\rho _{31}}{\Omega _{p}}=\frac{\Delta _{p}-\Delta _{c}+i\gamma }{%
\Omega _{c}^{2}/2-2\left( \Delta _{p}+i\Gamma /2\right) \left( \Delta
_{p}-\Delta _{c}+i\gamma \right) },  \label{rho31-steady}
\end{equation}%
which can be expanded as a power series of $\omega $:

\begin{equation}
\frac{\rho _{31}}{\Omega _{p}}=\sum b^{\left( i\right) }(\omega -\omega
_{p})^{i},  \label{rho31-expand}
\end{equation}%
where $\omega _{p}$ is the center frequency of the probe pulse. It has been
shown that for an input Gaussian probe pulse with a width $\tau _{0}$,
\begin{equation}
\Omega _{p,in}(t)=\Omega _{p0}e^{-t^{2}/\tau _{0}^{2}},  \label{Omegap-in}
\end{equation}%
the output probe pulse will take the form \cite{Kao}

\begin{equation}
\Omega _{p,out}(t)=\Omega _{p0}\left( \frac{\tau _{0}}{\tau _{w}}\right)
e^{-\beta -(t-\tau _{d})^{2}/\tau _{w}^{2}},  \label{Omegap-out}
\end{equation}%
where $\beta =L\eta \textit{Im}\left[ b^{\left( 0\right) }\right] $, $\tau
_{d}=L\eta \textit{Re}\left[ b^{\left( 1\right) }\right] $ and $\tau _{w}=%
\sqrt{\tau _{0}^{2}+4L\eta \textit{Im}\left[ b^{\left( 2\right) }\right] \text{
}}$, corresponding to the attenuation constant, delay time and the broadened
width of the output probe pulse, respectively. In deriving eq.($\ref%
{Omegap-out}$), we retain the power series in eq.($\ref{rho31-expand}$) merely
up to the quadratic terms and set $\Delta _{p}=\Delta _{c}$. Under the
approximations of strong coupling field, $\Omega _{c}^{2}\gg \gamma \Delta
_{p}$ and $\Omega _{c}^{2}\gg \gamma \Gamma $, $b^{(0)}$ and $b^{(2)}$ are
purely imaginary numbers and $b^{(1)}$ is a real number. The above
conditions for achieving eq.($\ref{Omegap-out}$) are well justified in the EIT
related experiments since the EIT bandwidth is proportional to $\Omega
_{c}^{2}/\Gamma $ and the frequency bandwidth of the probe pulse limits the
maximum value of $(\omega -\omega _{p})$. Note that eq.($\ref{Omegap-out}$) is
obtained based on the assumption that all atoms are kept fixed.
Realistically, the atoms move freely with a velocity distribution obeying
the Maxwell-Boltzmann statistics at finite temperatures. To reconcile these
two contradictory situations, eqs.($\ref{Bloch12}$)-($\ref{Bloch13}$) should be
understood as the motion equations seen by any observer fixed on the moving
atoms. To incorporate the Doppler shift into eq.($\ref{rho31-steady}$) in an
uncomplicated way, we shall assume that the probe pulse propagates along the
major-axis of the atomic cloud and the coupling field is applied with an
angle $\theta $ with respect to the major-axis. We also assume that the
center frequencies of coupling and probe fields are on two-photon resonance
in the laboratory frame, namely, $\Delta _{p}=\Delta _{c}$. Hence, for an
atom moving with a velocity $\mathbf{v}=\mathbf{v}_{\perp }\mathbf{+v}_{z}$%
\textbf{\ }in the non-relativistic limit,\textbf{\ }where\textbf{\ }$\mathbf{%
v}_{z}$\textbf{\ }is the velocity along the major-axis and\textbf{\ }$%
\mathbf{v}_{\perp }$\textbf{\ }is the velocity in the transverse direction,
it experiences the Doppler shifts so that\textbf{\ }$\Delta _{p}\rightarrow
\Delta _{p}-k_{p}v_{z}$\textbf{\ }and\textbf{\ }$\Delta _{c}\rightarrow
\Delta _{c}-\left( k_{c}v_{z}\cos \theta +k_{c}v_{\perp }\sin \theta \right)
$\textbf{\ }\cite{Jackson} . As a result, the condition of two-photon
resonance does not hold generally, and the two-photon detuning, $\delta
=\Delta _{p}-\Delta _{c}$, is maximized when $\theta =\pi $.

In the presence of Doppler shifts, all $b^{\left( i\right) }$ in eq.($\ref%
{rho31-expand}$) now turns out to be velocity-dependent, \textit{i.e.}, $%
b^{\left( i\right) }\rightarrow \tilde{b}^{\left( i\right) }(v_{z},v_{\perp
})$. Statistically, $b^{\left( i\right) }$ can be replaced by an effective
coefficient obtained by taking the ensemble average of $\tilde{b}^{\left(
i\right) }(v_{z},v_{\perp })$ over the Maxwell-Boltzmann velocity
distribution at a given temperature $T$,

\begin{equation}
\tilde{b}^{\left( i\right) }(\mathbf{v})\rightarrow \bar{b}^{\left( i\right)
}\left( T\right) =\frac{1}{\pi v_{s}^{2}}\int dv_{z}dv_{\perp }\tilde{b}%
^{\left( i\right) }(\mathbf{v})e^{-\mathbf{v}^{2}/v_{s}^{2}},  \label{bi-ave}
\end{equation}%
where $v_{s}=\sqrt{2k_{B}T/m}$ is the one-dimensional root-mean-square
velocity.\textbf{\ }It should be noted that the ensemble-averaging process
in eq.($\ref{bi-ave}$) is liable to the relaxation of $\rho _{21}$\ owing to
the atomic thermal motion. To be consistent, in each $\tilde{b}^{\left(
i\right) }(v)$, we let $\gamma _{T}=0$\ and keep $\gamma _{0}$ in $\gamma $\
while evaluating the coefficients\textbf{\ }$\bar{b}^{(i)}\left( T\right) $.
Our effective theory is formulated based on the above ensemble-averaging
process, in which the analytical forms of $\bar{\Omega}_{c}\left( T\right) $%
, $\bar{\eta}\left( T\right) $ and $\bar{\gamma}\left( T\right) $ are
derived. The basic idea relies on the presumption that, either the
stationary medium characterized by $\bar{\Omega}_{c}\left( T\right) $, $\bar{%
\eta}\left( T\right) $ and $\bar{\gamma}\left( T\right) $ or the
Doppler-broadened medium characterized by the temperature-independent $%
\Omega _{c}$, $\eta $ and $\gamma _{0}$, should lead to the same attenuation
constant, delay time and the broadened width of the output probe pulse,
namely, \textit{\ }%
\begin{equation}
\frac{\beta }{L}=\left. \eta \bar{b}^{(0)}(T)\right\vert _{\gamma
_{0},\Omega _{c}}=\left. \bar{\eta}b^{(0)}\right\vert _{\bar{\gamma},\bar{%
\Omega}_{c}},  \label{aq.1}
\end{equation}%
\begin{equation}
\frac{\tau _{d}(T)}{L}=\left. \eta \bar{b}^{(1)}(T)\right\vert _{\gamma
_{0},\Omega _{c}}=\left. \bar{\eta}b^{(1)}\right\vert _{\bar{\gamma},\bar{%
\Omega}_{c}},  \label{aq.2}
\end{equation}%
and
\begin{equation}
\frac{\tau _{w}^{2}(T)-\tau _{0}^{2}}{4L}=\left. \eta \bar{b}%
^{(2)}(T)\right\vert _{\gamma _{0},\Omega _{c}}=\left. \bar{\eta}%
b^{(2)}\right\vert _{\bar{\gamma},\bar{\Omega}_{c}}.  \label{aq.3}
\end{equation}%
We evaluate eqs.($\ref{aq.1})-(\ref{aq.3}$) by making use of eqs.($\ref%
{rho31-steady}$) and ($\ref{bi-ave}$), together with the assumption $\Delta
_{p}=\Delta _{c}=0$ for a stationary medium, and this yields%
\begin{eqnarray}
&&\int d^{2}v\frac{\left[ h\left( \mathbf{v},\theta \right) +i\gamma _{0}%
\right] e^{-v^{2}/v_{s}^{2}}}{\Omega _{c}^{2}/2-2\left[ h\left( \mathbf{v}%
,\theta \right) +i\gamma _{0}\right] G\left( v_{z}\right) }  \label{eq.1} \\
&=&\frac{\bar{\eta}\pi v_{s}^{2}}{\eta }\frac{\bar{\gamma}}{\bar{\Omega}%
_{c}^{2}/2+\Gamma \bar{\gamma}},  \nonumber
\end{eqnarray}

\begin{eqnarray}
&&\int d^{2}v\frac{\Omega _{c}^{2}/2+\left[ h\left( \mathbf{v},\theta
\right) +i\gamma _{0}\right] ^{2}e^{-v^{2}/v_{s}^{2}}}{\left[ \Omega
_{c}^{2}/2-2\left[ h\left( \mathbf{v},\theta \right) +i\gamma _{0}\right]
G\left( v_{z}\right) \right] ^{2}}  \nonumber \\
&=&\frac{\bar{\eta}\pi v_{s}^{2}}{\eta }\frac{\bar{\Omega}_{c}^{2}/2-2\bar{%
\gamma}^{2}}{\left( \bar{\Omega}_{c}^{2}/2+\Gamma \bar{\gamma}\right) ^{2}},
\label{eq.2}
\end{eqnarray}%
and

\begin{eqnarray}
&&\int d^{2}v\frac{4h^{3}\left( \mathbf{v},\theta \right) +\Omega _{c}^{2}%
\left[ 2h\left( \mathbf{v},\theta \right) +G(v_{z})\right] +g\left( \mathbf{v%
},\theta \right) }{\left[ \Omega _{c}^{2}/2-2\left[ h\left( \mathbf{v}%
,\theta \right) +i\gamma _{0}\right] G\left( v_{z}\right) \right] ^{3}}%
e^{-v^{2}/v_{s}^{2}}  \nonumber \\
&=&\frac{\bar{\eta}\pi v_{s}^{2}}{\eta }\frac{\Gamma \bar{\Omega}_{c}^{2}/2-4%
\bar{\gamma}^{3}+2\bar{\gamma}\bar{\Omega}_{c}^{2}}{\left( \bar{\Omega}%
_{c}^{2}/2+\Gamma \bar{\gamma}\right) ^{3}},  \label{eq.3}
\end{eqnarray}%
where $h\left( \mathbf{v},\theta \right) =kv_{\perp }\sin \theta
-kv_{z}\left( 1-\cos \theta \right) $, $G\left( v_{z}\right)
=-kv_{z}+i\Gamma /2$ and $g\left( \mathbf{v},\theta \right) =12\gamma
_{0}h\left( \mathbf{v},\theta \right) \left[ ih\left( \mathbf{v},\theta
\right) -\gamma _{0}\right] +2i\gamma _{0}\Omega _{c}^{2}$. To further
simplify the last three equations, we assume $\left\vert \sin \theta
\right\vert \ll 1$ together with the assumptions\textbf{\ }$\Omega
_{c}^{2}\gg \gamma _{0}\Gamma $\textbf{, }$\bar{\Omega}_{c}^{2}\gg \bar{%
\gamma}\Gamma $\textbf{, }$\Omega _{c}^{2}\gg \gamma _{0}\Delta _{p}$\textbf{%
, }$\Gamma \gg \bar{\gamma}$ and $k_{p}\approx k_{c}=k$ which are the
typical conditions for the EIT experiments\textbf{. }The condition\textbf{\ }%
$\left\vert \sin \theta \right\vert \ll 1$ is generally true, since so far
all EIT experiments have used either the nearly co-propagating ($\theta
\approx 0$) or counter-propagating ($\theta \approx \pi $) geometry\textbf{.
}In our experiments,\textbf{\ t}he central wavelength of the laser beam is $%
780$nm, $\gamma _{0}\simeq 5\times 10^{-4}\Gamma $\textbf{, }$\bar{\gamma}%
\simeq 2\times 10^{-3}\Gamma $,\textbf{\ }$\Delta _{p}=0$, and $0.6\Gamma
\leq \Omega _{c}\leq 0.9\Gamma $,\textbf{\ }where the spontaneous decay rate
of the excited state is $\Gamma =2\pi \times 5.9$\ MHz. Obviously, the
forementioned assumptions are valid and thus the effective parameters can
now be solved as

\begin{eqnarray}
&&\bar{\Omega}_{c}\left( T\right)  \label{OmegacT} \\
&=&\Omega _{c}\left\{ 1+4\left[ 2f\left( \theta \right) -\frac{4+5f\left(
\theta \right) }{\Gamma ^{2}}\Omega _{c}^{2}\right] x^{2}+\cdot \cdot \cdot
\right\} ,  \nonumber
\end{eqnarray}

\begin{eqnarray}
&&\bar{\eta}\left( T\right)  \label{etaT} \\
&=&\eta \left\{ 1+8\left[ f\left( \theta \right) -\frac{2+4f\left( \theta
\right) }{\Gamma ^{2}}\Omega _{c}^{2}\right] x^{2}+\cdot \cdot \cdot
\right\} ,  \nonumber
\end{eqnarray}%
and

\begin{equation}
\bar{\gamma}\left( T\right) =\gamma _{0}+4f\left( \theta \right) \frac{%
\Omega _{c}^{2}}{\Gamma }x^{2}+\cdot \cdot \cdot ,  \label{gammaT}
\end{equation}%
where\textbf{\ }$f\left( \theta \right) =1-\cos \theta $\textbf{\ }and%
\textbf{\ }$x=\sin \left( \theta /2\right) kv_{s}\Gamma /\sqrt{2}\Omega
_{c}^{2}$, which is the ratio of the effective Doppler width to that of the
EIT window.\ In eq.($\ref{gammaT}$), the leading term of $\bar{\gamma}\left(
T\right) $\ is the ground-state relaxation rate originating from the
intrinsic effects and the second term is resulted from the Doppler shifts%
\textbf{. }When the probe and coupling fields are nearly co-propagating,
i.e. $\theta \simeq 0$, the Doppler effect is considerably suppressed, and
this is why the EIT experiments at room temperatures are always studied in
the co-propagating geometry \cite{Budker} . On the contrary, when the probe
and coupling fields are in the nearly counter-propagating geometry, i.e. $%
\theta \simeq \pi $, the decoherence mechanism is dominated by Doppler
effects, and this is why the EIT-related experiments in the
counter-propagating geometry are always studied in cold atoms \cite{Liao}%
\textbf{.}

\begin{figure}[htbp]\begin{center}
\includegraphics[width=3in]{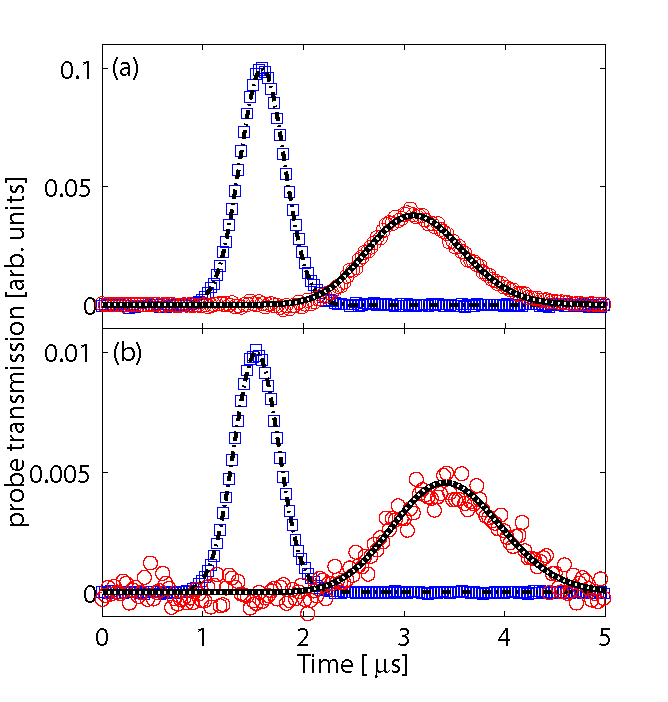}
\caption{(Color online) The
squares and circles are the experimentally measured intensities of the input
and output probe pulses, respectively. The input probe pulses are scaled
down by a factor of $0.1$ in (a) and $0.01$ in (b). The dashed-dotted lines
represent the numerical simulations of the input probe pulses. The black
lines represents the numerical results of the output probe pulse by solving
SL in the stationary EIT medium using the effective parameters $\left( \bar{%
\Omega}_{c},L\bar{\protect\eta},\bar{\protect\gamma}\right) =\left(
0.7\Gamma ,16.13\Gamma ,0.0143\Gamma \right) $ in (a) and $\left( 0.67\Gamma
,19.98\Gamma ,0.0227\Gamma \right) $ in (b). The gray dotted lines represent
the best fitting by using the scheme in \protect\cite{Wu} with the
parameters $\left( T,\Omega _{c},L\protect\eta \right) =\left( 205\protect%
\mu K,0.71\Gamma ,16.5\Gamma \right) $ in (a) and $\left( 305\protect\mu %
K,0.68\Gamma ,20.5\Gamma \right) $ in (b). Clearly, the black and gray
dotted lines completely merge in both (a) and (b).}
\label{sag}
\end{center}\end{figure}

\section{RESULTS AND DISCUSSIONS}

To demonstrate the availability of the current scheme, we use the
experimental data of SL as an input for eqs.($\ref{OmegacT}$)-($\ref{gammaT}$).
To highlight the effect of thermal motion of atoms, we consider the case
when the probe and coupling beams are arranged with $\theta \simeq \pi $
along the major axis of a cigar-shaped cloud of laser-cooled $^{87}$Rb atoms
\cite{Lin} .\ The experimental values of the coupling Rabi frequency, opacity
and relaxation rate of the ground-state coherence (denotes as $\Omega
_{c}^{\ast },L\eta ^{\ast }$ and $\gamma _{0}^{\ast }$) can be estimated by
the method described in \cite{Liao} . We apply the numerical scheme of \cite%
{Wu} , which considers all velocity groups of the atoms in the calculations,
to\ determine the temperature $T$\ of the atomic medium and get the best
fitting of $\Omega _{c}$\ and $L\eta $ as the gray dotted lines plotted in
Fig.2(a) and (b). To be generic, we consider two independent sets of
experimental data of SL with the estimated values $\left( \Omega _{c}^{\ast
},L\eta ^{\ast },\gamma _{0}^{\ast }\right) =\left( 0.687\Gamma ,15\Gamma
,0.0005\Gamma \right) $\ and $\left( 0.69\Gamma ,20.5\Gamma ,0.0005\Gamma
\right) $. The best fitting comes up correspondingly with $\left( T,\Omega
_{c},L\eta \right) =\left( 205\mu K,0.71\Gamma ,16.5\Gamma \right) $ and $%
\left( 305\mu K,0.68\Gamma ,20.5\Gamma \right) $. Inserting the above best
fitted parameters into eqs.($\ref{OmegacT}$)-($\ref{gammaT}$), the effective
parameters are found to be $\left( \bar{\Omega}_{c},L\bar{\eta},\bar{\gamma}%
\right) =\left( 0.7\Gamma ,16.13\Gamma ,0.0143\Gamma \right) $\ and $\left(
0.67\Gamma ,19.98\Gamma ,0.0227\Gamma \right) $. Using $\bar{\Omega}_{c},$\ $%
L\bar{\eta}$\ and $\bar{\gamma}$\ in eqs.($\ref{Bloch12}$)-($\ref{Maxwell}$), we
numerically calculate the output probe pulse as a function of time and find
it is in good agreement with the experimental data as the black lines
plotted in Fig.2(a) and (b). The validity of eqs.($\ref{OmegacT}$)-($\ref%
{gammaT}$) is hence demonstrated.

\begin{figure}[htbp]\begin{center}
\includegraphics[width=3in]{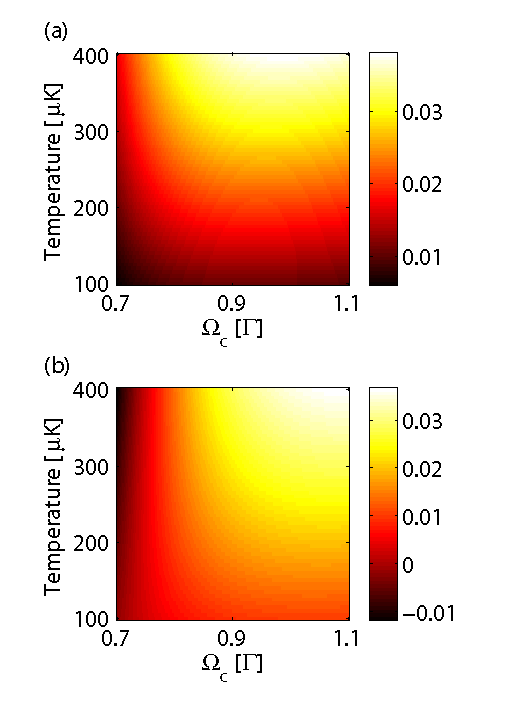}
\caption{(Color online) The fractional changes, $\protect\delta \bar{\Omega}%
_{c}=\left( \Omega _{c}-\bar{\Omega}_{c}\right) /\Omega _{c}$ and $\protect%
\delta \bar{\protect\eta}=\left( \protect\eta -\bar{\protect\eta}\right) /%
\protect\eta $ are plotted in (a) and (b) as functions of temperatures and $%
\Omega _{c}$ with $\protect\theta =\protect\pi $.}
\label{sag}
\end{center}\end{figure}

\begin{figure}[htbp]\begin{center}
\includegraphics[width=3in]{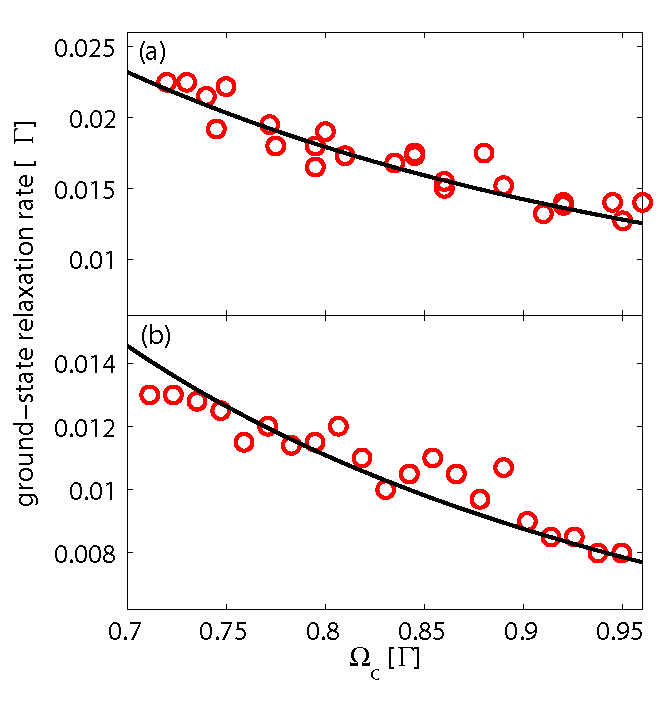}
\caption{(Color online) The comparisons between the measured $\protect\gamma $\ and the theoretically predicted $\bar{\protect\gamma}$ for $L\protect\eta %
^{\ast }$ $=20.5\Gamma $, $15\Gamma $\ in (a) and (b), respectively. The
circles are the measured data and the solid lines are the theoretical
predictions at $T=330\protect\mu $K, $200\protect\mu $K in (a) and (b). The
corresponding temperatures of the measured data in (a) range from $290%
\protect\mu $K to $370\protect\mu $K while those in (b) range from $175%
\protect\mu $K to $230\protect\mu $K.}
\label{sag}
\end{center}\end{figure}

The proximity of $\bar{\Omega}_{c}$ and $\bar{\eta}$ to their
zero-temperature counterparts can be examined by defining the fractional
changes, $\delta \bar{\Omega}_{c}=\left( \Omega _{c}-\bar{\Omega}_{c}\right)
/\Omega _{c}$ and $\delta \bar{\eta}=\left( \eta -\bar{\eta}\right) /\eta $,
which are shown in Fig.3 as functions of $\Omega _{c}$ and $T$ within the
experimentally accessible regime. Obviously, the small values of $\delta
\bar{\Omega}_{c}$ and $\delta \bar{\eta}$ shown in Fig.3 indicates that, $%
\Omega _{c}\approx \bar{\Omega}_{c}$ and $\eta \approx \bar{\eta}$, a
conclusion that is consistent with the perturbative results in eqs.($\ref%
{OmegacT}$)-($\ref{etaT}$). We thus expect that the atomic thermal motion
affects the ground-state relaxation rate mostly. In Fig.4, we compare the
measured ground-state relaxation rate\ with the theoretical prediction, eq.($%
\ref{gammaT}$). Here the measured relaxation rate is determined by fitting
the experimental data with the numerical results from eqs.(1)-(3). In Fig.4,
the theoretical predictions agree with the measured data very well and we
see that the averaged temperatures of our experimental setup of $L\eta
^{\ast }=15\Gamma $\ and $20.5\Gamma $ are found to be $T=200\mu $K and $%
330\mu $K, respectively.

\begin{figure}[htbp]\begin{center}
\includegraphics[width=3in]{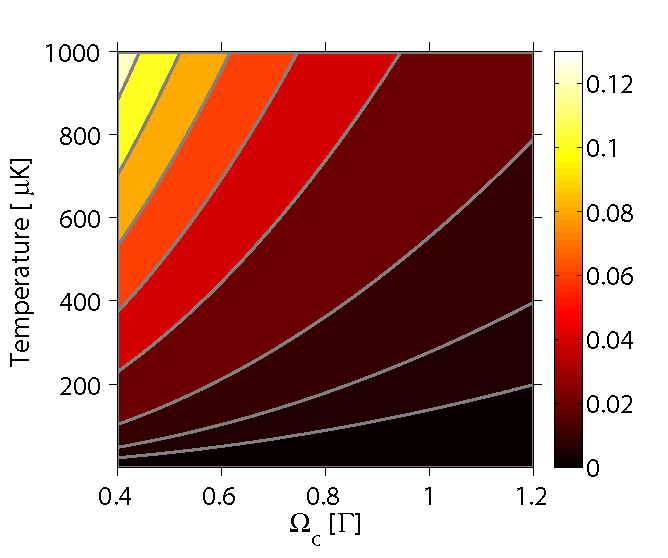}
\caption{(Color online) Contour plot of $\tilde{\protect\gamma}$ with $\protect\theta =\protect\pi $. The vertical axis ranges from $T=1\protect\mu $K to $1000%
\protect\mu $K and the horizontal axis ranges from $\Omega _{c}=0.4\Gamma $
to $1.2\Gamma .$ The values of the contour lines are from $0.12$ (the upper
left), $0.1$, $0.08$, $0.06$, $0.04$, $0.02$, $0.01$,$0.005$ (the lower
right).}
\label{sag}
\end{center}\end{figure}

\begin{figure}[htbp]\begin{center}
\includegraphics[width=3in]{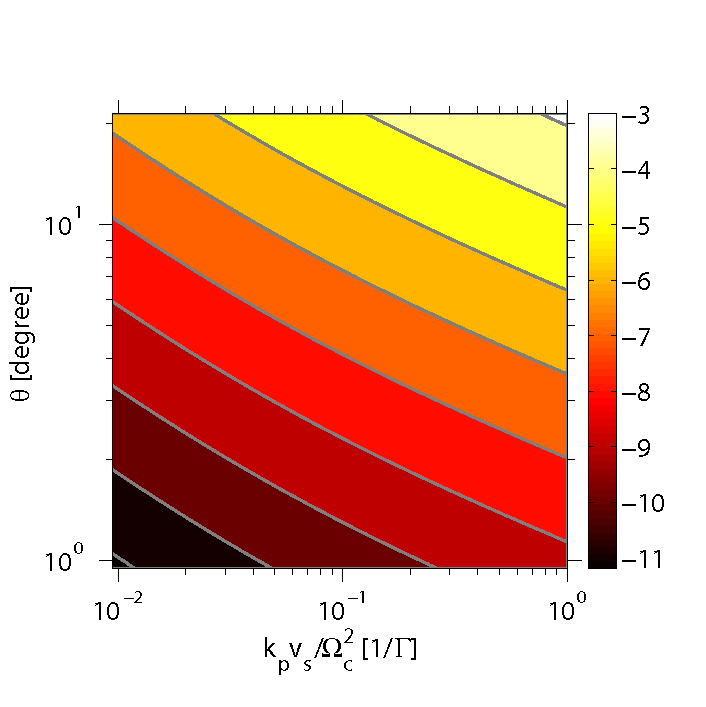}
\caption{(Color online) The plot of $\tilde{\protect\gamma}$ as
a function of $\protect\theta $ and $k_{p}v_{s}/\Omega _{c}^{2}$ in
logarithm scale. The vertical axis ranges from $\protect\theta =1^{\circ }$
to $21^{\circ }$ and the horizontal axis ranges from $k_{p}v_{s}/\Omega
_{c}^{2}=0.01\Gamma ^{-1}$ to $\Gamma ^{-1}.$ The values of the contour
lines is from $10^{-3}$ (the upper right) to $10^{-10}$ (the lower left)
with a spacing of one order of magnitude.}
\label{sag}
\end{center}\end{figure}

\begin{figure}[htbp]\begin{center}
\includegraphics[width=3in]{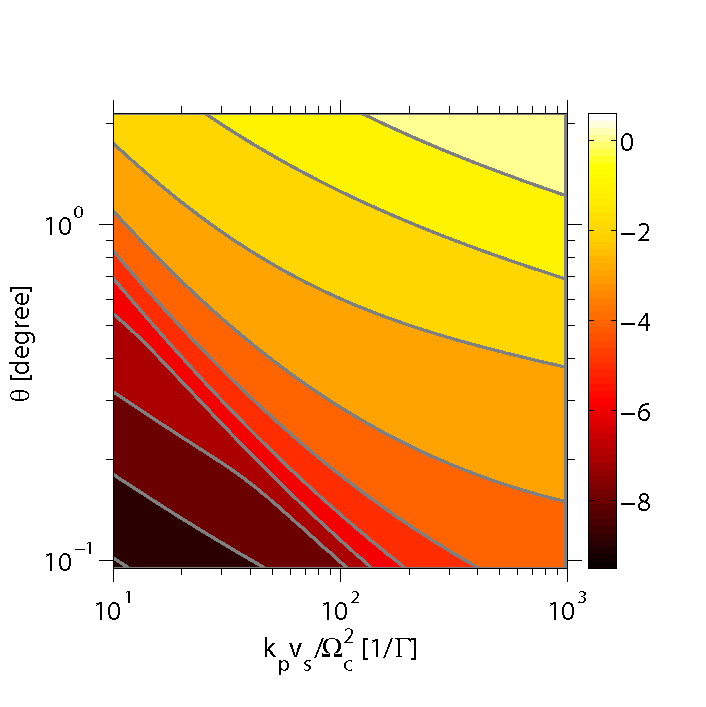}
\caption{(Color online) The plot of $\tilde{\protect\gamma}$ as a function of $%
\protect\theta $ and $k_{p}v_{s}/\Omega _{c}^{2}$. The vertical axis ranges
from $\protect\theta =0.1^{\circ }$ to $2.1^{\circ }$ and the horizontal
axis ranges from from $k_{p}v_{s}/\Omega _{c}^{2}=10\Gamma ^{-1}$ to $%
1000\Gamma ^{-1}.$ The values of the contour lines are from $10^{0}$ (the
upper right) to $10^{-8}$ (the lower left) with a spacing of one order of
magnitude.}\label{sag}
\end{center}\end{figure}
After demonstrating the validity of eq.($\ref{gammaT}$), we study how the
decoherence rate $\bar{\gamma}$ varies with $\Omega _{c},$ $T$ and $\theta $%
. Giving $\theta =\pi $, we first depict $\bar{\gamma}$ as a function of $%
\Omega _{c}$ and $T$ in Fig.5. Here we extend the ranges of $\Omega _{c}$
and $T$ so that the conditions of stationary light pulses and those using
the counter-propagating beam geometry with laser-cooled Rb atoms can be
included. We see that $\bar{\gamma}$ decays with increasing intensity of the
coupling field as shown in Fig.5. This is because the width of the
transparency window depends on the intensity of the coupling field, and an
increasing $\Omega _{c}$ can lower the absorption of the probe pulse, so
that $\bar{\gamma}$ would become smaller. The influence of the orientation
of the coupling beam and the effect of Doppler width on $\bar{\gamma}$ is
investigated by plotting $\bar{\gamma}$ as a function of $\theta $ and $%
k_{p}v_{s}/\Omega _{c}^{2}$ in Figs.6-7, which were obtained based on the
numerical calculations of eqs.($\ref{eq.1})-(\ref{eq.3}$). The conditions of
most experiments utilizing small separation angle between two propagating
beams in lase-cooled atoms are met in the ranges, $1^{\circ }\leq \theta $ $%
\leq $ $21^{\circ }$, $0.01\Gamma ^{-1}\leq k_{p}v_{s}/\Omega _{c}^{2}\leq $
$\Gamma ^{-1}$, as shown in Fig.6. On the other hand, those experiments
utilizing the nearly co-propagating beam geometry for hot atoms are mostly
achieved in the ranges $0.1^{\circ }\leq \theta $ $\leq $ $2.1^{\circ }$, $%
10\Gamma ^{-1}\leq k_{p}v_{s}/\Omega _{c}^{2}\leq $ $1000\Gamma ^{-1}$, as
shown in Fig.7. Taking our experimental conditions ($^{87}$Rb atoms and $%
\lambda =780$nm) as an example, if $T=100\mu $K then we have $%
k_{p}v_{s}/\Omega _{c}^{2}\approx 0.03\Gamma ^{-1}$. Figures 5-7 thus
provide useful information to determine the relaxation rate for those
EIT-related experiments in which the Doppler-broadening cannot be ignored.

\section{CONCLUDING REMARKS}

In conclusion, we have theoretically studied the effects of atomic thermal
motion on the dynamics of SL in a Doppler-broadened EIT medium. By extending
the results in \cite{Kao} , we have obtained the attenuation constant, the
delay time and the broadened width of the output probe field at finite
temperatures. We also have derived a set of parameters, $\bar{\Omega}_{c}$, $%
\bar{\eta}$ and $\bar{\gamma}$, which are temperature-dependent that serve
as effective thermal sources in the optical Bloch equation for a stationary
EIT medium. Our approach provides an approximate and convenient way to
determine the relaxation rate of SL instead of appealing to the more
complicated multiple-velocity calculations. This effective theory is not
only valid for the cold atomic medium but also applicable to the atomic
medium at room temperatures.

\textbf{ACKNOWLEDGEMENTS}

This work is supported by National Science Council, Taiwan, under Grant No.
98-2112-M-018-001-MY2 and No. 98-2628-M-007-004. S.-C. Gou acknowledges the
support from National Center for Theoretical Science.

\end{document}